\documentclass[10pt,twocolumn,letterpaper]{article}

\usepackage{cvpr}

\usepackage{graphicx}
\usepackage{booktabs}
\usepackage{multirow}
\usepackage{subcaption}
\usepackage[table]{xcolor}
\usepackage{stfloats}

\usepackage{tikz}
\usetikzlibrary{positioning,arrows.meta,calc}
\definecolor{cvprblue}{rgb}{0.21,0.49,0.74}
\definecolor{rows_in_table}{RGB}{40,160,160}
\usepackage[pagebackref,breaklinks,colorlinks,allcolors=cvprblue]{hyperref}

\def\paperID{8570} 
\def\confName{CVPR}
\def\confYear{2026}

\title{CADReasoner: Iterative Program Editing for CAD Reverse Engineering}

\author{Soslan Kabisov\textsuperscript{1}\\
\and
Vsevolod Kirichuk\textsuperscript{1}\\
\and
Andrey Volkov\textsuperscript{1}\\
\and
Marina Barannikov\textsuperscript{2} \\
\and
Gennadiy Savrasov\textsuperscript{1} \\
\and
Anton Konushin\textsuperscript{1} \\
\and
Andrey Kuznetsov\textsuperscript{134}
\\
\and
\centering
Dmitrii Zhemchuzhnikov\textsuperscript{1}\dag
\and 
\\ 
\textsuperscript{1} Lomonosov Moscow State University;  \textsuperscript{2} Université Paris Dauphine; \\
\textsuperscript{3} Innopolis University; 
\textsuperscript{4} FusionBrain Lab
}

\begin{document}
\maketitle

\begin{abstract}
\let\thefootnote\relax\footnotetext{\textsuperscript{\dag}Corresponding author: zhemchuzhnikovds@my.msu.ru}
Computer-Aided Design (CAD) powers modern engineering, yet producing high-quality parts 
still demands substantial expert effort. Many AI systems tackle CAD reverse engineering, 
but most are single-pass and miss fine geometric details. In contrast, human engineers 
compare the input shape with the reconstruction and iteratively make changes in the design
based on the remaining differences. Agent-based methods mimic this loop with frozen VLMs, 
but weak 3D grounding of current foundation models limits reliability and efficiency. We 
introduce CADReasoner, a model trained to iteratively refine its prediction based on the 
geometric discrepancy between the input and the predicted shape. The model outputs a runnable CadQuery Python program, feeding its rendered mesh back to the next step.
CADReasoner fuses multi-view renders and point clouds as complementary modalities. To 
bridge the realism gap, we propose a scan-simulation protocol applied during both training 
and evaluation. Across DeepCAD, Fusion 360, and MCB benchmarks, CADReasoner attains state-of-the-art results on clean and scan-sim tracks. Code and the model are available at
\href{https://github.com/zhemdi/CADReasoner}{GitHub}
and \href{https://huggingface.co/kulibinai/cadreasoner}{Hugging Face model}.
\end{abstract}

\section{Introduction}

\begin{figure*}[t]
  \centering
  \includegraphics[width=\linewidth]{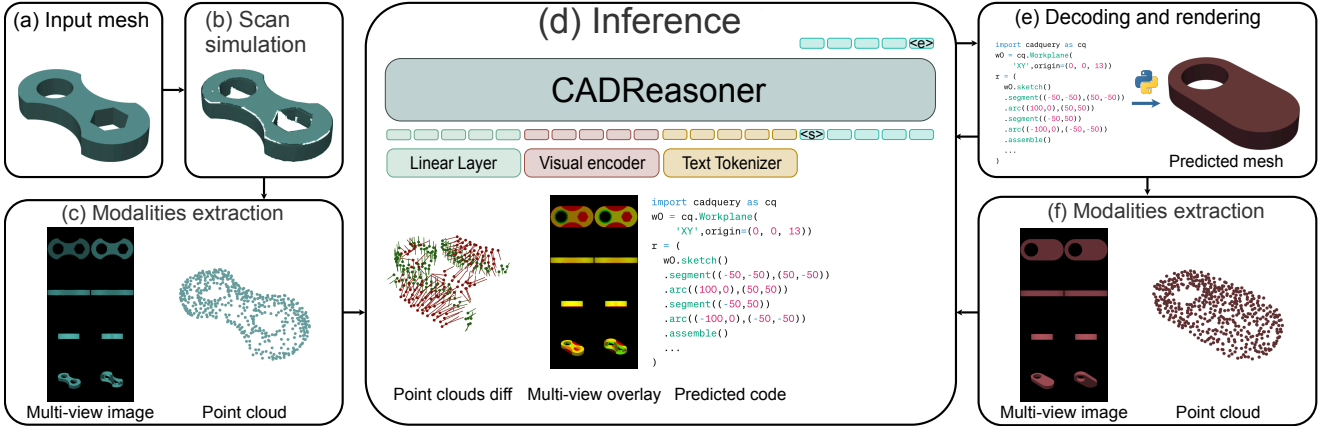} 
\caption{\textbf{CADReasoner: scan-to-CAD self-editing loop.}
(a–c) From a (simulated) scanned mesh we extract target evidence: multi-view images and a point cloud.
(d) CADReasoner consumes this evidence together with the current program $C^{t-1}$ and autoregressively predicts an updated \texttt{CadQuery} script $C^{t}$.
(e–f) The script is executed to render a new mesh, from which modalities are re-extracted, closing the geometry-driven self-correction loop.}
  \label{fig:cadreasoner_arch}
\end{figure*}

Recovering precise CAD models from raw 3D scan data is a fundamental yet challenging problem in computer vision and graphics ~\cite{avetisyan2019scan2cad,mallis2023cc3d,lin2025survey}. Scanned 3D data (e.g., point clouds or meshes obtained via LiDAR or RGB-D sensors) typically capture real-world objects with noise and missing regions, whereas CAD models are clean, parametric representations with exact surfaces and structures. Bridging this gap requires algorithms that can interpret imperfect scan data and produce accurate, realistic CAD models. Existing approaches in AI-based CAD reconstruction often either rely on single-shot predictions or simple alignment techniques that lack the ability to iteratively refine errors. As a result, they may output CAD models that deviate from the true geometry or miss important details, especially when faced with noisy or incomplete scans.

A key limitation of many prior methods is the absence of an effective mechanism for iterative inference and self-correction. In contrast, human engineers naturally refine designs by repeatedly comparing the evolving part against the target and correcting discrepancies.
Some recent systems mimic this loop with frozen VLM/LLM backbones~\cite{alrashedygenerating,li2025seek}, but their 3D grounding is limited and the backbone does not learn from its own errors. Other works incorporate geometry in the RL fine-tuning stage \cite{kolodiazhnyi2025cadrille,guan2025cad-coder-grpo,wang2025gaco}, typically as scalar rewards, which provides a coarse signal that departs from how practitioners reason; moreover, emerging analyses suggest  RL fine-tuning largely reweights existing behaviors rather than injecting new task knowledge
~\cite{yue2025does,shao2025spurious}. By contrast, we put geometry at the center of supervised fine tuning: the editor is trained to refine its own prediction from cross-modal discrepancy evidence (multi-view overlays and nearest-surface offsets) together with the previous program, aligning supervision with real-world, iterative reverse-engineering practice. In this view, self-editing becomes a minimal, learnable implementation of the 3D \textbf{reasoning} loop an engineer performs when inspecting and revising a design.

A second challenge is \emph{realism}. Public datasets are synthetic: B-Rep objects come from CAD corpora, and inputs are rendered views or point clouds (PCs) sampled from B-Reps rather than raw scans~\cite{willis2021fusion360,wu2021deepcad,rukhovich2024cad-recode,doris2025cad-coder,chen2025cadcrafter}. Real reverse engineering starts from physical parts: scans exhibit noise, occlusions and missing patches, extra or missing holes, and mild mis-registration. We therefore introduce a scan-simulation pipeline and apply it during training and evaluation, so the editor learns on scan-like inputs and is evaluated under the same conditions.

To summarize, our contributions are as follows:
\begin{itemize}
    \item We introduce \textbf{CADReasoner}, a novel framework for AI-based CAD reconstruction that performs iterative inference and \textbf{self-correction}, enabling progressively refined alignment between CAD models and the input.
    \item We bring \textbf{scan realism}  to CAD reconstruction by applying a scan-simulation regime for both training and evaluation, better reflecting real reverse-engineering pipelines.
    \item We demonstrate through extensive experiments that CADReasoner achieves state-of-the-art results on \textbf{DeepCAD}, \textbf{Fusion360}, and \textbf{MCB} outperforming existing methods in terms of reconstruction accuracy and realism. Our approach shows especially strong improvements in challenging scenario with scan defects.
\end{itemize}

\section{Related Work}

\paragraph{CAD generation.}
Prior work on CAD generation
can be categorized
by target representation: \emph{CSG trees}, \emph{B-reps}, and \emph{program (sequence) models}. CSG composes primitives with boolean ops, but struggles with intricate, real-world geometry and diverges from how designers actually build parts~\cite{du2018inversecsg,ellis2019write,friedrich2019optimizing,kania2020ucsg-net,nandi2018functional,ren2021csg-stump,tian2019learning,yu2022capri-net,yu2023d2csg}. B-reps encode faces,edges and vertices with strict topological validity, making generation brittle and subsequent editing cumbersome~\cite{guo2022complexgen,jayaraman2022solidgen,lambourne2021brepnet,li2019supervised,li2025caddreamer,liu2024split,liu2024point2cad,sharma2020parsenet,smirnov2021learning,wang2022neural,wang2020pie-net,xu2024brepgen}. In contrast, program or sequence formulations (sketch–extrude, feature histories, or Python code) align with industrial workflows and preserve editability~\cite{wu2021deepcad, lambourne2022prismcad, ren2022extrudenet, xu2022skexgen, xu2023hnc-cad, zhang2024flexcad, badagabettu2024query2cad, chen2024img2cad, khan2024cad-signet, khan2024text2cad, ma2024cad-diffuser, mallis2024cad-assistant, li2025cad-llama, doris2025cad-coder, he2025cad-coder, yuan2025cad-editor, wang2025cad-gpt}. Among these, recent methods that emit concise, executable \texttt{CadQuery} \cite{cadquery} programs achieve state-of-the-art performance on multiple benchmarks~\cite{rukhovich2024cad-recode,kolodiazhnyi2025cadrille,wang2025gaco}, so we likewise adopt \texttt{CadQuery} as our target representation.

\paragraph{Multi-view and point-cloud conditioning.}
AI-based CAD reverse engineering has largely been \emph{single-modality}. Methods conditioned on \emph{point clouds}—e.g., DeepCAD, CAD-Recode, CAD-SIGNet—predict programs from surfaces sampled on B-Reps\cite{wu2021deepcad,rukhovich2024cad-recode,khan2024cad-signet}. Image-conditioned approaches—CADCrafter, CADCoder, GACO-CAD—operate on rendered  \emph{views}\cite{chen2025cadcrafter,doris2025cad-coder,wang2025gaco}. Multimodal models are only now emerging: \textit{cadrille} accepts either views or clouds, while CAD-MLLM fuses both\cite{kolodiazhnyi2025cadrille,xu2024cad-mllm}. We regard these as complementary shape representations—views expose global inconsistencies , clouds provide metric nearest-surface distances—and use them as feedback to drive \emph{iterative} program editing.

\begin{table*}[t]
\centering
\caption{\textbf{Image-only, clean surfaces, greedy decoding.} Median CD$\downarrow$ ($\times 10^{3}$), mean IoU$\uparrow$ (\%), IR$\downarrow$ (\%). Best-so-far at $t\in\{1,5\}$. \textbf{CADReasoner} delivers \emph{superior performance}: it achieves the best CD/IoU/IR on \textbf{DeepCAD} and \textbf{Fusion360} already by $t{=}5$, and on \textbf{MCB} it markedly improves IoU/IR over \textit{cadrille} while remaining competitive in CD.}
\label{tab:img_main}
\setlength{\tabcolsep}{8pt} 
\begin{tabular}{lccc ccc ccc}
\toprule
& \multicolumn{3}{c}{DeepCAD} & \multicolumn{3}{c}{Fusion360} & \multicolumn{3}{c}{MCB} \\
\cmidrule(lr){2-4}\cmidrule(lr){5-7}\cmidrule(lr){8-10}
Method & CD & IoU & IR & CD & IoU & IR & CD & IoU & IR \\
\midrule
CADCrafter            & --   & --   & --   & 0.26 & --   & 3.6 & --   & --   & --   \\
cadrille          & 0.18 & 86.1 & 1.5  & 0.20 & 77.6 & 3.2 & \textbf{1.21} & 40.2 & 8.6 \\
\midrule
\rowcolor{rows_in_table!15}
CADReasoner ($t{=}1$) & 0.17 & 88.7 & 1.4  & 0.16 & 81.1 & 3.5 & 3.5 & 39.6 & 21.9 \\
\rowcolor{rows_in_table!29}
CADReasoner ($t{=}5$) & \textbf{0.16} & \textbf{90.0} & \textbf{0.5} &
                         \textbf{0.15} & \textbf{83.1} & \textbf{0.7} &
                         1.67 & \textbf{42.7} & \textbf{4.0} \\
\bottomrule
\end{tabular}
\end{table*}

\begin{figure*}[t]
  \centering
  \includegraphics[width=\textwidth]{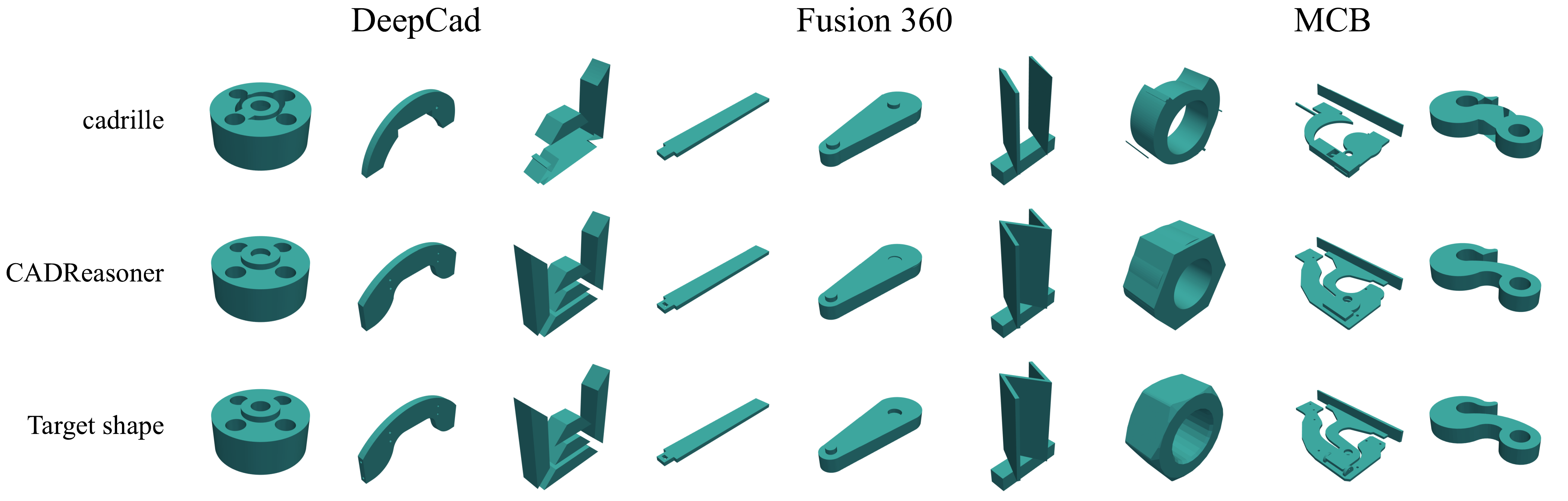} 
  \caption{\textbf{Qualitative comparisons on clean surfaces (multi-view, greedy).}
Nine examples (three each from \textbf{DeepCAD}, \textbf{Fusion360}, \textbf{MCB}). 
Each panel shows \emph{Target}, \emph{cadrille-SFT}, and \emph{CADReasoner} at $t{=}5$ (best-so-far).
Across datasets, \textbf{CADReasoner} produces \emph{more accurate reconstructions}, recovering small features and correct topology that \emph{cadrille} often misses or distorts.}
  \label{fig:qual_img}
\end{figure*}

\paragraph{Iterative refinement and code editing.}
Outside CAD, iterative self-correction for reasoning uses self-feedback or verbal RL without weight updates (Self-Refine; Reflexion) and RL with unit-test signals (CodeRL), but none provides geometric supervision\cite{madaan2023self,shinn2023reflexion,le2022coderl}. 
Jones et al.\ propose \emph{Learning to Edit Visual Programs with Self-Supervision}, which iteratively refines visual programs via predicted local edit operations from rendered feedback\cite{jones2024learning}. 
In contrast, we focus on CAD reverse engineering with executable \texttt{CadQuery}, use geometry discrepancy encodings (multi-view overlays + nearest-surface offsets), support images and point clouds, and refine by directly rewriting the program with a single trained model (no explicit edit vocabulary/\texttt{findEdits} or separate edit network).
Within CAD, agentic pipelines keep the backbone frozen: Seek-CAD runs a training-free loop with DeepSeek-R1 using visual feedback from Gemini-2.0; VideoCAD targets UI-interaction learning and uses DINOv2 descriptors for quality validation; CADCodeVerify prompts a VLM to validate render–target similarity and suggest fixes, again without a trained self-editing refiner\cite{li2025seek,man2025videocad,alrashedygenerating}. 
\emph{cadrille} introduced large-scale SFT followed by RL driven by geometry-based rewards. We instead use geometrical feedback to train the model to refine its predictions on the SFT stage.

\paragraph{Scan-based reverse engineering.}
In practical reverse engineering the source is a physical part; we acquire a 3D scan whose geometry is imperfect (noise, missing patches, extra or missing holes, mild mis-registration). A large body of work consumes point clouds and predicts structured CAD: CSG or primitive assemblies (InverseCSG; UCSG-Net; CAPRI-Net; D2CSG), B-rep surfaces/topology (Point2CAD), or full CAD programs (DeepCAD; CAD-Recode; CAD-SIGNet)\cite{du2018inversecsg,kania2020ucsg-net,yu2022capri-net,yu2023d2csg,liu2024point2cad,wu2021deepcad,rukhovich2024cad-recode,khan2024cad-signet}. Although close to the scan setting (a scan is also a point set), these inputs are typically sampled from the ideal 3d model surfaces rather than captured by sensors. Some program-induction works inject Gaussian jitter on points (e.g., CAD-Recode; \textit{cadrille}), but this does not capture characteristic scan defects \cite{rukhovich2024cad-recode,kolodiazhnyi2025cadrille}. We therefore introduce a scan-simulation pipeline that explicitly models such defects and apply it during training and evaluation, so the editor learns to operate on scan-like inputs and is evaluated under the same conditions. A notable step toward realism is the CC3D dataset, which virtually scans CAD models with a proprietary pipeline to produce paired scan–surface data exhibiting protrusions, smoothing, and missing regions~\cite{mallis2023cc3d}.

\section{Method}

\begin{table*}[t]
\centering
\caption{\textbf{Point-cloud, clean surfaces, greedy decoding.} Median CD$\downarrow$ ($\times 10^{3}$), mean IoU$\uparrow$ (\%), IR$\downarrow$ (\%). Best-so-far at $t\in\{1,5\}$. \textbf{CADReasoner} achieves \emph{superior performance} to \textit{cadrille} across all three benchmarks by $t{=}5$ (lower CD/IR, higher IoU). A dash (–) denotes numbers not reported by the original work.}
\label{tab:pc_main}
\begin{tabular}{lccccccccc}
\toprule
& \multicolumn{3}{c}{DeepCAD} & \multicolumn{3}{c}{Fusion360} & \multicolumn{3}{c}{MCB} \\
Method & CD & IoU & IR & CD & IoU & IR & CD & IoU & IR \\
\midrule
CAD-SIGNet & 0.28 & 77.6 & 0.9& 0.56 & 65.6 & 1.6 & - & - & - \\
MiCADangelo & 0.20 & 80.6 & 2.6 & 0.48 & 68.7 & 3.2 & - & - & - \\
CAD-Recode &  0.18&  87.1& 3.1 & 0.19  & 79.1 & 5.0 & 0.75 & 44.3 & 12.5 \\
cadrille & 0.18 & 87.1 & 2.1 & 0.19 & 79.8 & 2.8 & 0.82 & 45.7 & 4.0 \\
\midrule
\rowcolor{rows_in_table!15}
CADReasoner ($t{=}1$) &  0.21& 86.1& 1.5 & 0.22 & 77.3 & 3.2 & 1.03 & 41.3  & 16.2 \\
\rowcolor{rows_in_table!29}
CADReasoner ($t{=}5$) & \textbf{0.17}  &  \textbf{89.2} & \textbf{0.2} & \textbf{0.16} & \textbf{83.1} & \textbf{0.7} & \textbf{0.52} & \textbf{49.1} & \textbf{3.1} \\
\bottomrule
\end{tabular}
\end{table*}

\subsection{Problem Setup}
Let $T$ be the target shape (triangulated B-rep or scan) and let $\mathcal{R}$ maps a parametric CAD program $C$ (in our case, a runnable \texttt{CadQuery}~\citep{cadquery} Python script) to a mesh $S=\mathcal{R}(C)$. The goal is to deliver a program $C^\star$:
\[
C^\star \in \arg\min_{C\in\mathcal{C}_{\text{valid}}} D\!\big(T,\mathcal{R}(C)\big),
\]
where $D$ is the task-native geometric discrepancy 
and
$\mathcal{C}_{\text{valid}}$ denotes the conceptual set of all valid scripts that compile and yield a non-degenerate solid.

Rather than one-shot prediction, we exploit that geometry feedback is intrinsic:
given any candidate $C$, we can execute $\mathcal{R}(C)$ and compare to $T$.
We therefore cast reverse engineering as closed-loop program editing. At iteration $t$,
\[
C^{t} = f_{\theta}\!\big(E(T,S^{t-1}),\,C^{t-1}\big), \qquad S^{t-1}=\mathcal{R}(C^{t-1}),
\]
where $E(T,S^{t-1})$ encodes the pair (multi-view overlays and point-set features) and
$f_{\theta}$ is a learned editor that emits an updated \texttt{CadQuery} script. After each step we
retain the best-so-far program by $D$ and stop when improvement falls below a threshold or a step budget is reached.

\subsection{CADReasoner Architecture}
CADReasoner is implemented using a vision–language model (VLM). It accepts three inputs per iteration: (i) multi-view image obtained by overlaying the
target and the previous render (G/R channels), encoded by a shared visual backbone; (ii) a
point-cloud branch where per-point features 
are projected
by a linear layer and aggregated by a set encoder; and (iii) the previous program $C^{t-1}$,
tokenized as text. Tokens from geometry branches form a geometry memory that conditions the
code decoder. The decoder generates the updated script $C^{t}$ that is executed to produce $S^{t}$, which is fed to the model in the next step. The CADReasoner architecture is illustrated by Figure \ref{fig:cadreasoner_arch}.

\begin{table*}[t]
\centering
\caption{\textbf{Cross-modality (images + PCs), clean surfaces, greedy decoding.} Median CD$\downarrow$ ($\times 10^{3}$), mean IoU$\uparrow$ (\%), IR$\downarrow$ (\%). Best-so-far at $t\in\{1,5\}$. \textbf{CADReasoner} fuses complementary cues—global silhouette/topology from multi-view images and metric nearest-surface offsets from point sets. On \textbf{DeepCAD}/\textbf{Fusion360}, cross-modal is on par with image-only (images already suffice), while on \textbf{MCB} fusion is \emph{decisive}: it lowers IR and lifts IoU relative to single branches, approaching the PC branch’s best CD. Overall, cross-modal CADReasoner attains the lowest invalidity at $t{=}5$ across datasets, with competitive or best CD/IoU.}

\label{tab:fusion_main}
\begin{tabular}{lccccccccc}
\toprule
& \multicolumn{3}{c}{DeepCAD} & \multicolumn{3}{c}{Fusion360} & \multicolumn{3}{c}{MCB} \\
Method & CD & IoU & IR & CD & IoU & IR & CD & IoU & IR \\
\midrule
\rowcolor{rows_in_table!15}
CADReasoner - img ($t{=}1$) & \textbf{0.17} & \textbf{88.5} & 1.6  & \textbf{0.17} & \textbf{82.3} & 4.0 & 2.48 & 39.6 & 21.9 \\
\rowcolor{rows_in_table!15}
CADReasoner - pc ($t{=}1$) &  0.21& 86.1& 1.5 & 0.22 & 77.3 & \textbf{3.2} & \textbf{1.03} & \textbf{41.3}  & 16.2 \\
\rowcolor{rows_in_table!15}
CADReasoner - cross-modal ($t{=}1$) &  \textbf{0.17}& 88.3& \textbf{1.5} & \textbf{0.17} & \textbf{82.4} & 4.0 & 1.32 & 39.7  & \textbf{15.4} \\

\midrule
\rowcolor{rows_in_table!29}
CADReasoner - img ($t{=}5$) & \textbf{0.16} & \textbf{89.4} & 0.6 & \textbf{0.16} & \textbf{84.2} & 0.9 & 1.67 & 42.7 & 4.0 \\
\rowcolor{rows_in_table!29}
CADReasoner - pc  ($t{=}5$) & 0.17  & 89.2 & 0.2 & 0.18 & 83.1 & 0.7 & \textbf{0.52} & \textbf{49.1} & 3.1 \\
\rowcolor{rows_in_table!29}
CADReasoner - cross-modal  ($t{=}5$) & \textbf{0.16}  & \textbf{89.4} & \textbf{0.3} & \textbf{0.16} & \textbf{84.1} & \textbf{0.8} & 0.71 & 48.2 & \textbf{2.8} \\

\bottomrule
\end{tabular}
\end{table*}

\subsection{Geometry Encodings}
\paragraph{Multi-view image.}
We use $8$ canonical views: six orthographic projections and two isometric. Since all meshes in the CAD-Recode dataset lie within the cube \[\{ (x, y, z) \in \mathbb{R}^3 \mid -100 \le x, y, z \le 100 \}\]
for each projection plane an image of this region with the rendered mesh is captured and represented with a resolution of $238 \times 238$ pixels. For orthographic views, the depth of the mesh points along the view axis is encoded by the color intensity. Additionally, the $-Z$, $+Y$, $+X$ projections are mirrored from left to right so that, for each projection plane, the axes are aligned in the same direction on both images. The projection images are combined into 2 columns and 4 rows
. 
For each view $v$ the image of the target mesh $T$ and the predicted mesh $S^{t-1}$ are stored in the Green and Red channels of an RGB image respectively: $I_v=[\,R_v,\,G_v,\,0\,]$, where $G_v=D_v(\Pi_v(T))$ and $R_v=D_v(\Pi_v(S^{t-1}))$, $\Pi_v$ – orthographic projection and $D_v$ – depth to color mapping. Visually, \emph{green} marks geometry present in $T$, \emph{red} marks geometry present in $S^{t-1}$, and their overlap appears \emph{yellow}. An example of multi-view visualization is shown in the supplementary material (Figure ~\ref{fig:multi-view}).


\paragraph{Point clouds.}
At iteration $t$ we form two clouds: $P_T=\{p_i\}_{i=1}^{N_T}$ sampled on the target $T$ and
$P_S=\{q_j\}_{j=1}^{N_S}$ sampled on the previous render $S^{t-1}$. Both are expressed in a common
frame: we normalize $T$ to $[-1,1]^3$ and isotropically scale the prediction domain $[-100,100]^3$
by $1/100$ before computing cross-shape relations. Prior to constructing $P_T$ and $P_S$, we draw a dense set of 
$\sim\!30\,000$ surface samples from each mesh.
Let $\mathrm{NN}_X(y)$ be the nearest point to $y$
on surface $X$. 
We select 128 points from each mesh by choosing those with the largest distances to the other mesh.
For each sample we attach a cross-shape offset
\[
\Delta p_i=\mathrm{NN}_{S^{t-1}}(p_i)-p_i,\qquad
\Delta q_j=\mathrm{NN}_{T}(q_j)-q_j,
\]
so every point is encoded by $(x,y,z,\Delta x,\Delta y,\Delta z)$; 
i.e., a point together with the vector toward the nearest location on the opposite
mesh.
the set encoder consumes
$\{(p_i,\Delta p_i)\}\cup\{(q_j,\Delta q_j)\}$ to produce $z_{\text{pc}}$. Offsets act as a local,
metric error signal-magnitude measures discrepancy; direction indicates how the prediction
should move.

\textit{Initialization ($t{=}1$).}
With no prior render we adopt a null prediction at the origin while keeping the interface identical
to later steps. We sample $P_T$ on $T$ and set $\Delta p_i=-p_i$. For the second cloud we take
$P_S=\{\mathbf{0}\}^{N_S}$ and pair each $q_j=\mathbf{0}$ with a distinct target sample via a
permutation $\pi$, setting $\Delta q_j=p_{\pi(j)}$. This yields informative, nonidentical offsets at
$t{=}1$ without special tokens or modality changes.


\subsection{Scan-Simulation}
\label{sec:scan-sim}
To simulate the process of real camera-based scanning we developed a virtual 3D scanning algorithm. We uniformly sample 100{,}000 surface points from the CAD model while preserving the original
normals. A virtual camera then moves along a spherical trajectory around the object, and visible points are extracted at
five distinct viewpoints using visibility criteria~\cite{mehra2010visibility}. All visible points are merged, and the combined point cloud is converted into a mesh using
Screened Poisson Surface Reconstruction~\cite{kazhdan2013screened}. As a result, typical scanning artifacts – such as incomplete surface coverage and smoothed edges – are preserved. Finally, random holes are added to simulate missing regions produced by scanning markers.


\begin{table*}[t]
\centering
\caption{\textbf{Stochastic sampling ($N{=}5$ per step).} Median CD$\downarrow$ ($\times 10^{3}$), mean IoU$\uparrow$ (\%), IR$\downarrow$ (\%) at $t\in\{1,5\}$; best-so-far selection per step under an \emph{equal} sampling budget. \textbf{CADReasoner} achieves \emph{state-of-the-art} performance in this regime: by $t{=}5$ it attains the best (or tied-best) CD on \textbf{DeepCAD} ($0.15$) and \textbf{Fusion360} ($0.13$), zero invalidity (IR$=0$) across modalities/datasets, and competitive IoU, surpassing both supervised and RL-finetuned baselines overall. On \textbf{MCB}, the PC branch reaches the lowest CD ($0.40$) with IR$=0$.}
\label{tab:sampling}
\begin{tabular}{lcccccccccccc}
\toprule
& \multicolumn{3}{c}{DeepCAD} & \multicolumn{3}{c}{Fusion360} & \multicolumn{3}{c}{MCB} \\
Method &  CD & IoU & IR &  CD & IoU & IR &  CD & IoU & IR\\

\midrule
CAD-Recode (pc) &0.17 &89.2&0.5&0.16&80.4&0.9&0.62&46&6.5\\
cadrille-SFT (img) &0.17 &87.7&0.2&0.18&78.9&1.0&0.90&41.8&6.6\\
cadrille-SFT (pc) &0.17 &89.1&0.4&0.17&80.8&0.6&0.60&47&6.1\\
cadrille-RL (img) &0.16 &92.8&0.0&0.16&85.5&0.17&0.70&48.7&2.6\\
cadrille-RL (pc) &0.16 &92.1&0.1&0.15&\textbf{87.4}&0.5&\textbf{0.40}&\textbf{55.2}&2.9\\
\midrule
\rowcolor{rows_in_table!15}
CADReasoner (img) ($t{=}1$)   &0.16&90.7&0.1&0.15&83.4&\textbf{0.0}&1.47&42.7&\textbf{0.0}\\
\rowcolor{rows_in_table!15}
CADReasoner (pc) ($t{=}1$)   &0.2&87.5&\textbf{0.0}&0.22&79.3&0.1&0.84&45.6&\textbf{0.0}\\
\rowcolor{rows_in_table!15}
CADReasoner (cross-modal) ($t{=}1$)&0.16&91.1&0.1&0.15&84.0&0.2&1.47&42.1&1.3\\
\midrule
\rowcolor{rows_in_table!29}
CADReasoner (img) ($t{=}5$)   & 0.15&92.6&\textbf{0.0}&\textbf{0.13}&87.0&\textbf{0.0}&0.54&49.9&\textbf{0.0}\\
\rowcolor{rows_in_table!29}
CADReasoner (pc) ($t{=}5$)   &0.16 &92.1&\textbf{0.0}&0.2&85.5&\textbf{0.0}&\textbf{0.40}&53.4&\textbf{0.0}\\
\rowcolor{rows_in_table!29}
CADReasoner (cross-modal) ($t{=}5$)&\textbf{0.15} &\textbf{92.9}&\textbf{0.0}&\textbf{0.13}&86.5&\textbf{0.0}&0.55&48.8&\textbf{0.0}\\
\bottomrule
\end{tabular}
\end{table*}

\subsection{Training}
\paragraph{Backbone and single-model setup.}
Our editor $f_\theta$ is instantiated as \textbf{Qwen2-VL 2B}. We fine-tune this single model via SFT on the CAD-Recode corpus (with the curriculum below). Importantly, we use one shared checkpoint for all iterations: the same model performs initialization ($t{=}1$) and subsequent refinement steps ($t{>}1$).
\paragraph{SFT with curriculum.}

Training $t{=}1$ on the full CAD-Recode corpus can lead to \emph{train-set} overfitting (CAD-Recode is relatively simple), producing overly strong one-shot predictions on seen shapes and leaving little signal to learn refinement. 
We therefore use disjoint data splits and on-policy rollouts so refinement examples come from \emph{imperfect} predicted programs on \emph{unseen} targets.

\textit{Stage A (seed, $t{=}1$ on $\mathcal{D}_1$).} We train the editor by supervised cross-entropy to predict the ground-truth program $C^\star$ from $E(T,\varnothing)$ (no prior render/code). At the end model reliably emits valid but yet inaccurate \texttt{CadQuery} scripts.

\textit{Rollout B.} Using the Stage-A checkpoint, we generate one-step predictions $C^1$ for every $T\!\in\!\mathcal{D}_2$, rendering $S^1=\mathcal{R}(C^1)$.

\textit{Stage B (learn to refine).} We now train on a mixture of $(t{=}1)$ and $(t{=}2)$ examples: for $t{=}1$, inputs are $E(T,\varnothing)$; for $t{=}2$, inputs are $E(T,S^1)$ with context $C^1$, and the target remains $C^\star$. Thus the model learns to (i) generate from scratch and (ii) refine its own first-step predictions.

\textit{Rollout C.} With the Stage-B checkpoint, we produce two-step rollouts on $\mathcal{D}_3$: $C^1\!\to\!S^1$ and then $C^2=f_\theta(E(T,S^1),C^1)$ with $S^2=\mathcal{R}(C^2)$.

\textit{Stage C (extend horizon).} We train on $t\in\{1,2,3\}$, conditioning $t{=}2$ on $(C^1,S^1)$ and $t{=}3$ on $(C^2,S^2)$, always supervising toward $C^\star$. Batches sample $t$ uniformly.

This curriculum keeps supervision on-policy: each refinement step sees the distribution of errors produced by the current model, preventing collapse to trivial $t{=}1$ behavior and yielding a robust editor that can both initialize and self-correct.


\subsection{Inference and Selection}
\label{sec:inference}
We use two decoding regimes.

\paragraph{Greedy (single path).}
At each iteration $t$ the editor decodes deterministically (argmax), compiles $C^{t}$, renders
$S^{t}=\mathcal{R}(C^{t})$, and updates the best-so-far program
$C^{\le t}=\arg\min_{i\le t}D\!\big(T,\mathcal{R}(C^{i})\big)$. Complexity is $O(s)$ renders for
$s$ steps.

\paragraph{Stochastic, geometry-guided beam.}
We run a stochastic beam with geometry pruning. At $t{=}1$, sample $N$ candidates
$\{C^{1}_{k}\}_{k=1}^{N}$, compile and render them, rank by the primary
metric $D$ (invalid generations are discarded), and keep the top $N$ survivors. For $t{>}1$, from each
survivor generate $N$ children (at most $N^{2}$ candidates per step), render and score all, then retain
the top $N$ for the next iteration. We always maintain and report the best-so-far program
over all evaluated candidates. The total number of renders is
\(
N \;+\; (s-1)N^{2}.
\)

\paragraph{Evaluation and scoring protocol (both regimes).}
At evaluation step $s$ we always report the \emph{best-so-far} program
\[
C^{\le s} \;=\; \arg\min_{i\le s} D\!\big(T_{\mathrm{eval}},\,\mathcal{R}(C^{i})\big),
\]
i.e., metrics at step $s$ are computed on $\min_{i\le s} D$ (not only on $i{=}s$).

For clean experiments $T_{\mathrm{eval}}=T_{\text{clean}}$ (input is a triangulated B-Rep ).
For the scan track we never use the oracle clean mesh: \emph{all} ranking, pruning, tie-breaking,
and reporting use the scan evidence as the target, i.e., $T_{\mathrm{eval}}=T_{\text{scan}}$.
Thus candidate selection in the stochastic beam and the greedy updates are driven by the scan,
ensuring the procedure does not access ground truth unavailable at inference.

\section{Experiments}

\subsection{Datasets} 

\textbf{Training set.} We use the procedurally generated \textbf{CAD-Recode} corpus~\cite{rukhovich2024cad-recode}
($\sim1$M \texttt{CadQuery} programs), an order of magnitude larger than DeepCAD, to pretrain the editor. 
\\
\textbf{Evaluation sets.}
\begin{enumerate}
    \item \textbf{DeepCAD}~\cite{wu2021deepcad}  test split: 8{,}046 parts authored by users of Onshape. \\ 
    \item \textbf{Fusion360}~\cite{willis2021fusion360} provides a smaller (1{,}725 samples) but geometrically richer benchmark of real design sequences. \\
    \item \textbf{MCB (5k subset)}~\cite{sangpil2020large} We evaluate on a curated 5k-shape subset of \emph{MCB}, stratified by ISO part categories with approximately uniform sampling per category (equal quotas per ISO class). Parts are filtered for watertightness and moderate geometric complexity, and the splits preserve the category balance. \\
\end{enumerate}
For all evaluation datasets we prepare two tracks: clean (original meshes) and scan-sim produced by our simulator; the same simulator is optionally applied at train time.

\begin{figure*}[t]
\centering
\includegraphics[width=0.8\textwidth]{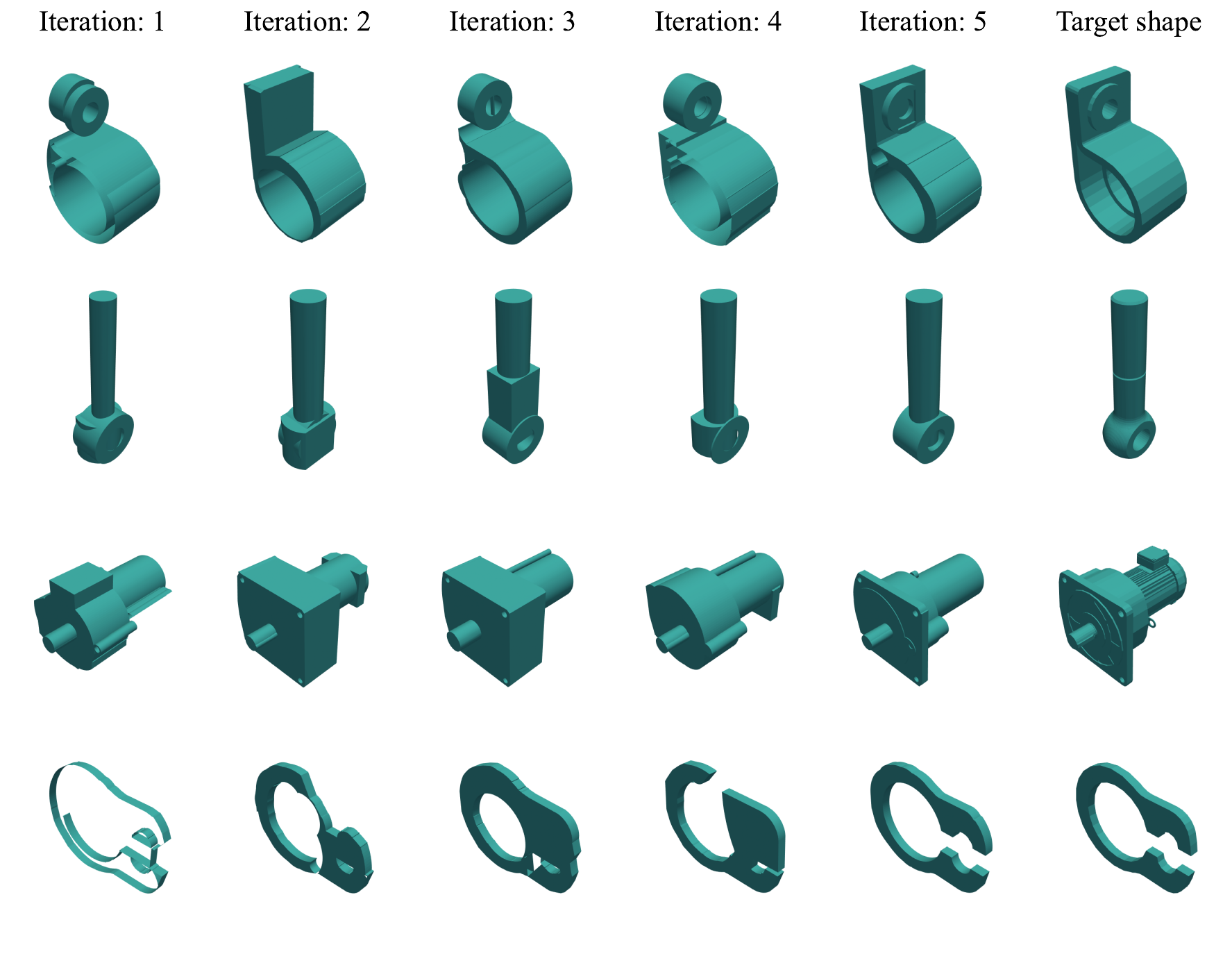}
\caption{\textbf{Qualitative evolution on clean surfaces (sampling).} CADReasoner’s iterative self-editing under the sampling regime ($N{=}5$ per step). Columns show $t{=}1-5$ and the target. Predictions progressively repair topology and recover small features.}
\label{fig:quals}
\end{figure*}

\subsection{Metrics}
All meshes are rigid-aligned and normalized to $[0,1]^3$. Following \cite{rukhovich2024cad-recode,kolodiazhnyi2025cadrille} we report:
(i) \textbf{Chamfer Distance (CD)} on $8{,}192$ vs $8{,}192$ points (CD values are multiplied by 10\textsuperscript{3}); 
(ii) \textbf{volumetric IoU} (\%); 
(iii) \textbf{Invalid Rate (IR)}: percentage of generations that fail to compile or yield a degenerate solid. 
To reduce bias from invalids, we report median CD instead of mean.

\textit{Selection vs reporting.} On the clean track, both selection and reporting use the clean target. 
On the scan-sim track, candidate selection uses CD against the scan to avoid oracle access, while final reported metrics are computed against the corresponding clean surface, so scores are comparable across tracks.

\subsection{Results}

\paragraph{Multi-view images (clean, greedy).}
We evaluate CADReasoner \emph{without} an RL stage and therefore compare to supervised-only baselines (notably \emph{cadrille-SFT}). The question is whether iterative self-editing improves quality as the step count $t$ increases. We start from the simplest setting: clean meshes, multi-view overlays, and greedy decoding (no sampling). The editor runs for $t$ steps; we report best-so-far at $t\!\in\!\{1,5\}$. 
Table~\ref{tab:img_main} shows that on \textbf{DeepCAD} and \textbf{Fusion360} CADReasoner already beats the baselines at $t\!=\!1$ and improves further at $t\!=\!5$ (lower CD/IR, higher IoU), likely helped by our $8$-view overlays (vs.\ $4$ in \textit{cadrille}). On \textbf{MCB}, $t\!=\!1$ trails due to higher geometric complexity, but five iterations yield large gains that surpass \textit{cadrille-SFT} on IoU and IR. Figure~\ref{fig:qual_img} (nine cases; three per dataset) illustrates more accurate reconstructions: CADReasoner recovers small features and correct topology that \textit{cadrille-SFT} often misses.

\paragraph{Point clouds (clean, greedy).}
We then train and evaluate the point-set branch alone, comparing to CAD-SIGNet, MiCADangelo, CAD-Recode, and \emph{cadrille-SFT} (Table~\ref{tab:pc_main}). At $t\!=\!1$ our model is on par or slightly below the strongest SFT baselines (e.g., CD $0.21$ vs.\ $0.18$ on DeepCAD), consistent with the weaker global cues in a single point set. After self-editing to $t\!=\!5$, CADReasoner exceeds \textit{cadrille-SFT} across all three benchmarks and sharply reduces invalidity-e.g., MCB IR $16.2 \!\to\! 3.1$-indicating that nearest-surface offsets guide effective code repairs while tightening local geometry.
\paragraph{Cross-modal (images + point clouds, clean, greedy).}
Because test inputs originate from meshes, both evidences are available; we therefore evaluate cross-modality of multi-view overlays and point sets. Table~\ref{tab:fusion_main} contrasts CADReasoner with image-only, PC-only, and fused inputs, reporting best-so-far at $t\!\in\!\{1,5\}$. On \textbf{DeepCAD} and \textbf{Fusion360}, cross-modal learning largely \emph{matches} image-only on CD/IoU and consistently \emph{reduces} IR,
indicating that metric offsets from PCs help the editor avoid brittle code paths even when silhouette cues already suffice. On \textbf{MCB}, it improves robustness, achieving the \emph{lowest} IR at both $t\!=\!1$ and $t\!=\!5$, while PC-only attains the best CD (and slightly higher IoU) at $t\!=\!5$. 

\begin{table*}[t]
\centering
\caption{\textbf{Simulated-scan results (sampling, $N{=}5$) — \underline{Main Table}.} Selection uses CD against the \emph{scan}; final metrics are computed against the corresponding \emph{clean} surfaces. \textbf{CADReasoner} achieves \emph{state-of-the-art} robustness under scan artifacts on \textbf{DeepCAD}, \textbf{Fusion360}, and \textbf{MCB}, outperforming both supervised and RL-finetuned baselines (e.g., \textit{cadrille-RL}) at $t{=}5$ while maintaining IR$=0$ across modalities. Best CD is obtained by the PC branch on MCB ($0.49$) and by the cross-modal branch on DeepCAD/Fusion360 ($0.16/0.15$).}
\label{tab:scan}
\begin{tabular}{lcccccccccccc}
\toprule
& \multicolumn{3}{c}{DeepCAD} & \multicolumn{3}{c}{Fusion360} & \multicolumn{3}{c}{MCB} \\
Method &  CD & IoU & IR &  CD & IoU & IR &  CD & IoU & IR\\
\midrule
CAD-Recode (pc) &0.28 &71.2&1.2&0.27&65.0&0.8&1.03&38.9&7.3&\\
cadrille-SFT (img) &1.14 &45.5&2.5&1.10&38.7&2.4&2.01&24.3&9.2&\\
cadrille-SFT (pc)  &0.26 &75.0&0.6&0.25&69&0.9&0.88&49.3&6.4&\\
cadrille-RL (img)  &0.47 &64.4&0.3&0.49&58.3&0.2&1.44&36.8&2.7&\\
cadrille-RL (pc)   &0.21 &83.7&0.2&0.19&79.2&0.2&0.55&50.7&2.7&\\
\midrule
\rowcolor{rows_in_table!15}
CADReasoner (img) ($t{=}1$)   &0.17 &88.0&0.1&0.16&81.9&0.1&1.04&46.4&0.2&\\
\rowcolor{rows_in_table!16}
CADReasoner (pc) ($t{=}1$)   &0.20 &85.8&0.0&0.22&78.3&0.1&0.72&47.4&\textbf{0.0}&\\
\rowcolor{rows_in_table!15}
CADReasoner (cross-modal) ($t{=}1$)&0.17 &88.2&0.0&0.16&81.2&0.0&1.10&46.4&\textbf{0.0}&\\
\midrule
\rowcolor{rows_in_table!29}
CADReasoner (img) ($t{=}5$)   &0.17 &\textbf{89.2}&\textbf{0.0}&\textbf{0.15}&\textbf{83.4}&\textbf{0.0}&0.55&50.5&\textbf{0.0}&\\
\rowcolor{rows_in_table!29}
CADReasoner (pc) ($t{=}5$)   &0.18 &86.6&\textbf{0.0}&0.18&79.8&\textbf{0.0}&\textbf{0.49}&50.8&\textbf{0.0}&\\
\rowcolor{rows_in_table!29}
CADReasoner (cross-modal) ($t{=}5$)&\textbf{0.16} &88.8&\textbf{0.0}&\textbf{0.15}&83.3&\textbf{0.0}&0.56&\textbf{51.3}&\textbf{0.0}&\\
\bottomrule
\end{tabular}
\end{table*}

\subsection{Sampling}
Greedy decoding can be suboptimal after SFT because the training objective is next-token likelihood (conditioned on discrepancy evidence), not direct geometric matching. In practice, multiple plausible edits may exist at a step, and the greedy choice can miss higher-quality trajectories. We therefore evaluate stochastic decoding: at each step we draw $N{=}5$ candidates per parent and retain the best-by-CD candidate (best-so-far over steps), under a fixed inference budget.

Table~\ref{tab:sampling} compares CADReasoner (image / PC / cross-modal) to CAD-Recode (PC), \emph{cadrille-SFT (img/pc)}, and \emph{cadrille-RL (img/pc)} with the same sampling budget. Despite using SFT only, CADReasoner + sampling already matches or exceeds \textit{cadrille-RL} on \textbf{DeepCAD} and \textbf{Fusion360} at $t{=}1$, and widens the margin by $t{=}5$. On \textbf{MCB}, CADReasoner(PC) reaches the lowest CD and IR at $t{=}5$. Overall, iterative self-editing with stochastic evaluation reliably outperforms RL-fine-tuned baselines under equal compute, by exploring diverse edits and selecting trajectories that sharpen small features while repairing invalid code. Figure~\ref{fig:quals} illustrates some sampled trajectories over iterations. The samples are from the MCB dataset, which contains higher-complexity shapes compared to the training dataset. In such cases, refinement is not necessarily monotonic: the model may temporarily add or remove elements and the best prediction can occur several iterations after the initial refinement.

\subsection{Results on scans}
We repeat the sampling experiment on the scan-sim track
(Sec.~\S\ref{sec:scan-sim})
.
Selection uses CD computed against the scan; final reporting uses the corresponding
clean surface 
so scores remain comparable across tracks.
Table~\ref{tab:scan} shows that CAD-Recode and \emph{cadrille} degrade substantially under
scan artifacts, while CADReasoner-trained with scan simulation-retains performance close to
the clean setting. On \textbf{DeepCAD} and \textbf{Fusion360}, CADReasoner (image/cross-modal) attains near-clean
CD with IR$\,{=}\,0$ by $t\!=\!5$; on \textbf{MCB}, the PC branch achieves the best CD (and IR$\,{=}\,0$),
and cross-modality gives the highest IoU. These results indicate that the editor learns to use scan-specific
signals (occlusions, missing patches) without overfitting to artifacts. Additional results on scans are provided in the supplementary material (Sec. B), including results on the CC3D dataset ~\cite{CC3D}.

\section{Conclusion}

We introduced \textbf{CADReasoner}, a closed-loop editor that reconstructs parametric CAD by \emph{iteratively} self-editing its CadQuery program using explicit geometry feedback. Treating multi-view overlays and point sets as complementary evidence, the model conditions on the target and the previous render to propose improved code at step $t$, with best-so-far selection by CD. To bridge realism, we proposed a \emph{scan-simulation} regime applied during training and evaluation. On \textbf{DeepCAD}, \textbf{Fusion360}, and \textbf{MCB}, CADReasoner establishes state-of-the-art results on clean and scan-sim tracks, substantially reducing CD and IR while increasing IoU, and recovering fine geometric details over iterations.

\clearpage
\section*{Acknowledgements}

This work was supported by the Ministry of Economic Development of the Russian Federation in accordance with the subsidy agreement (agreement identifier 000000C313925P4H0002; grant No 139-15-2025-012).

\small
\bibliographystyle{ieeenat_fullname}
\bibliography{main}

\clearpage
\normalsize
\clearpage
\setcounter{page}{1}
\setcounter{section}{0}   
\maketitlesupplementary
\renewcommand{\thesection}{\Alph{section}}

\section{Multi-view visualization}

\begin{figure}[!h]
  \centering
  \includegraphics[width=0.5\linewidth, keepaspectratio]{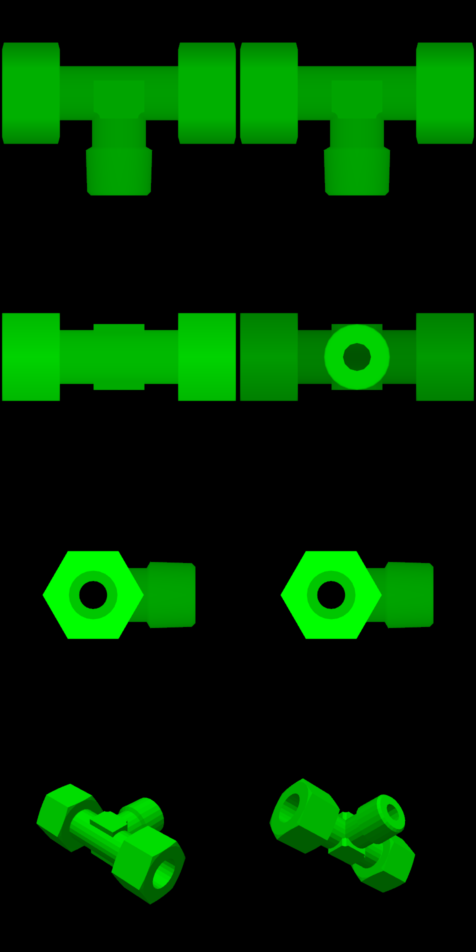}
  \caption{\textbf{Multi-view image of target shape.} We render eight fixed views—six orthographic (\(\pm X,\pm Y,\pm Z\)) and two isometric—using parallel projection and flat shading; intensity encodes depth. The views are arranged in a \(2\times4\) grid with consistent axis orientation. 
  }
  \label{fig:multi-view}
\end{figure}

Figure \ref{fig:multi-view} shows the input image for the first step, where the predicted shape is absent. For subsequent steps, the predicted shape is rendered in a similar way and incorporated into the red channel of the image.

\label{sec:multi-view}

\section{More results on scans}
\label{sec:more-scans}

To probe generalization across domains, Table~\ref{tab:train-scan-test-clean} compares four train/test
combinations. Training on scans improves transfer in both directions: compared to training on
clean data, it reduces CD on clean test and boosts
IoU. In short, scan-simulation acts as a robust curriculum that
yields models strong on scans and at least as strong on clean surfaces.

\begin{table}[!ht]
\centering
\caption{\textbf{Train–test cross-generalization under scan-simulation (CADReasoner, greedy decoding, iteration $t=5$, MCB subset:  1000 samples.} 
Selection uses scan CD when \emph{Test: scan}; final metrics are always computed on the corresponding \emph{clean} surface for comparability.
Median CD$\downarrow$ ($\times10^{3}$), mean IoU$\uparrow$ (\%), IR$\downarrow$ (\%).}
\label{tab:train-scan-test-clean}
\begin{tabular}{llccc}
\toprule
Train set & Test set & CD $\downarrow$ & IoU $\uparrow$ & IR $\downarrow$ \\
\midrule
clean & clean & 1.43 & 41.0 & 0.0 \\
\rowcolor{rows_in_table!29}
clean & scan  & 4.42 & 17.8 & 0.1 \\
scan  & clean & \textbf{1.01} & \textbf{47.4} & 0.0 \\
\rowcolor{rows_in_table!29}
scan  & scan  & 1.14 & 46.4 & 0.0 \\
\bottomrule
\end{tabular}
\end{table}

As reported in Table \ref{tab:cc3d}, we additionally evaluate our method on the CC3D dataset – a dataset of CAD models that includes both clean meshes and versions that have been virtually scanned using a proprietary 3D scanning pipeline, exhibiting realistic scan-like artifacts that differ from those generated by our simulation procedure. We present experimental results on the image modality because, unlike point clouds, which consist of only 256 points, images capture all scan defects and more clearly demonstrate the benefits of scan-sim training. Training with our scan simulation significantly reduces performance degradation on scanned data.

\clearpage

\begin{table*}[t]
\centering
\caption{\textbf{Results on CC3D dataset} (sampling, $N{=}5$). Final metrics are computed against the corresponding \emph{clean} surfaces.}
\label{tab:cc3d}

\begin{tabular}{lcccccccccccc}
\toprule
& \multicolumn{3}{c}{CC3D-clean} & \multicolumn{3}{c}{CC3D-scan}  & \multicolumn{2}{c}{$\Delta$ \%}\\
Method &  CD & IoU & IR &  CD & IoU & IR &  CD & IoU\\
\midrule
%
%
cadrille-SFT   &0.88 &54.9&8.4&1.15&48.4&10.2&+30.7&-11.8\\
%
%
cadrille-RL   &0.57 &65.0&0.1&0.67&60.3&0.1&+17.5&-7.2\\
\midrule
%
%
Our (w/o scan-sim) ($t$=1)  &0.81 &58.1&0.1&0.96&53.6&0.1&+18.5&-7.7\\
%
%
Our (w/o scan-sim) ($t$=5)   &0.39 &64.2&0.0&0.44&60.1&0.0&+12.8&-6.4\\
\midrule
%
%
Our (w/ scan-sim) ($t$=1)  &0.53 &62.0&0.1&0.55&61.2&0.1&\textbf{+3.8}&\textbf{-1.3}\\
%
%
Our (w/ scan-sim) ($t$=5)   &0.34 &66.9&0.0&0.37&66.7&0.0&\textbf{+8.8}&\textbf{-0.3}\\
\bottomrule
\end{tabular}
\end{table*}

\vspace*{\fill}


\end{document}